# Open Mass Spectrometry Search Algorithm

*Lewis Y. Geer*[*,†], *Sanford P. Markey*[‡], *Jeffrey A. Kowalak*[‡], *Lukas Wagner*[†], *Ming Xu*[†], *Dawn M. Maynard*[‡], *Xiaoyu Yang*[‡], *Wenyao Shi*[‡], *Stephen H. Bryant*[†]

National Center for Biotechnology Information, National Library of Medicine, National Institutes of Health, Bethesda, Maryland 20894, and the Laboratory of Neurotoxicology, National Institute of Mental Health, National Institutes of Health, Bethesda, Maryland 20892

Running Head: OMSSA

[*] To whom correspondence should be addressed. E-mail: lewisg@mail.nih.gov

[†] National Center for Biotechnology Information.

[‡] Laboratory of Neurotoxicology.

Large numbers of MS/MS peptide spectra generated in proteomics experiments require efficient, sensitive and specific algorithms for peptide identification. In the Open Mass Spectrometry Search Algorithm [OMSSA], specificity is calculated by a classic probability score using an explicit model for matching experimental spectra to sequences. At default thresholds, OMSSA matches more spectra from a standard protein cocktail than a comparable algorithm. OMSSA is designed to be faster than published algorithms in searching large MS/MS datasets.

# Introduction

High throughput proteomics [1,2] involves the analysis of thousands of peptide spectra derived from biological samples. These spectra can be identified by four general types of algorithms: de novo calling of the sequence directly from the spectrum [3-5], the use of unambiguous "peptide sequence tags" derived from spectra that are used to search known sequences [6-8], cross-correlation methods that correlate experimental spectra with theoretical spectra [9,10], and probability-based matching that calculates a score based on the statistical significance of a match between an observed peptide fragment and those calculated from a sequence search library [11-15]. The Open Mass Spectrometry Search Algorithm (OMSSA) is an algorithm of the last type and is unique in its use of classical hypothesis testing based on an explicit model of matching statistics, the type of statistical model used in BLAST [16]. Due to the large numbers of spectra generated in high throughput proteomics, manual interpretation is impractical, so it is essential that the matches from these algorithms are scored with a threshold or thresholds that allow few false positives. Probability-based matching allows such thresholds to be set in terms of the number of false positives allowed, leading to the development of several statistical scoring algorithms [17-24] in addition to existing probability based search algorithms. OMSSA is an attempt to create a fast search algorithm whose results are scored using a classical statistical model using assumptions taken directly from the experimental setup and allowing for experimental noise.

OMSSA takes experimental ms/ms spectra, filters noise peaks, extracts *m/z* values, and then compares these *m/z* values to calculated *m/z* values derived from peptides produced by an *in silico* digestion of a protein sequence library. The theoretical peptides must have a mass within a user specified tolerance of the precursor mass. The resulting search hits are then statistically scored.

To validate OMSSA, a standard protein cocktail at several concentrations was analyzed by an automated ion-trap 2D LC-MS/MS system and the resulting spectra evaluated by OMSSA and Mascot [11], a commonly used probability-based search algorithm. The search results from both algorithms were then compared.

**Experimental section**

The protein standards used in this paper were obtained from Sigma-Aldrich Corp. St. Louis, MO. A standard cocktail was prepared and composed of horse myoglobin (Accession P02188), bovine serum albumin (Accession P02769), chicken lysozyme (Accession P00698), and bovine carbonic anhydrase II (Accession AAO85140) using 10 fmol and 100 fmol aliquots. The yeast lysate used in this study consists of *S. cerevisiae* proteins sequentially extracted from the original 5g pellet using a modified MudPIT procedure [25]. The cocktail was denatured, reduced, alkylated and digested with endoproteinase Lys-C followed by trypsin. The digest (20 ug) was analyzed by an automated 2D LC-MS/MS system [26]. This system is composed of Shimadzu LC-VP series components directly coupled to a ThermoFinnigan LCQ Classic ion trap mass spectrometer. Peptide samples were injected via an auto sampler and loaded onto a strong cation exchange column. Peptides were then eluted in 6 steps of increasing ionic strength and captured onto 6 individual peptide cap traps. The contents of the cap traps were sequentially eluted onto a reverse phase column and the peptides were subsequently separated and eluted directly into the ESI ion trap mass spectrometer. The 100 fmol protein standard dataset contained 486 spectra and the 10 fmol protein standard dataset contained 134 spectra, all of which were encoded into dta file format using the LCQ_DTA.exe utility. MS/MS spectra were acquired using a "big five" scan function, i.e., iterative acquisition of a full scan mass spectrum from 400-1800 *m/z* followed by sequential CID tandem mass spectra of the five most abundant precursor ions from the previous full scan mass spectrum. No attempt was made to determine the charge state of the precursor ions. Without knowledge of precursor charge state, the LCQ-DTA.exe utility assigns the charge state as 1+ if no product ions are observed above the *m/z* value of the precursor ion. Otherwise, the LCQ_DTA.exe utility calculates relative molecular weight ($M_r$) for both the 2+ and 3+ charge states.

The OMSSA algorithm is written in C++ using the standard NCBI C and C++ toolkits. The use of the NCBI toolkits allows the compilation and use of OMSSA on most major operating systems, such as Linux, Windows, MacOS, and Solaris. The source code for OMSSA is in the public domain and is part of the NCBI C++ toolkit, available at ftp://ftp.ncbi.nih.gov/toolbox/ncbi_tools++/CURRENT/. By using the C toolkit, OMSSA is able to use sequence libraries created for BLAST [16], simplifying the creation or acquisition of sequence search libraries. Executables for OMSSA may be found at ftp://ftp.ncbi.nlm.nih.gov/pub/lewisg/omssa.

OMSSA takes as input two file formats: dta file format and a generic file format. The generic file format is described in the OMSSA source code. We intend to add additional file input formats to support additional instruments.

The version of Mascot used for this paper was 1.8. Mascot was run on a cluster of four two-CPU machines running Red Hat Linux 7, where three of the machines were computational nodes. The CPU's in this cluster were 1.4 GHz AMD MP/Athlon's with 1 Gb main memory shared on each machine. OMSSA was run on one CPU of a two-CPU Linux 2.4 machine using 1.26 GHz Pentium III processors with 2 Gb of main memory.

Searches in both OMSSA and Mascot were done with the following settings: tryptic cleavage, precursor mass tolerance of ±2 Da, product mass tolerance of ±0.8 Da, carbamidomethyl cysteine as a fixed modification, methionine oxide as a variable modification, and 1 missed cleavage allowed.

## Results and Discussion

The general flow of the algorithm is shown in Figure 1. The individual steps are detailed below.

**Charge Determination.** MS/MS spectra often do not include experimentally measured precursor charges for reasons of sensitivity. However, determining the possible precursor charge states of spectrum is necessary in OMSSA because it selects theoretical peptides from the sequence search library by comparison to the neutral mass of the precursor. To do this, OMSSA determines whether the precursor is charge 1+ or not by counting the number of peaks above the precursor *m/z*. If more than 95% of the peaks are below the precursor *m/z* value, the spectrum is determined to be 1+, otherwise the spectrum is determined to be charge $\geq$ 2+, and is searched twice, assuming that the precursor ion is 2+ or 3+.

**Noise filter.** Experimental spectra can contain a significant amount of noise that can result in random matches, requiring the algorithm to eliminate some noise peaks without deleting signal peaks. Fortunately, it is not necessary to delete all noise peaks as the search statistics take noise into account. The OMSSA noise filter has several steps. The first step is to delete background peaks whose intensity is below a percentage of the maximum intensity peak. By default, OMSSA cuts peaks below 2.5% of the maximum intensity, although this value is user adjustable and is modified dynamically in the last part of the algorithm, as detailed in a later section. The second step is to delete any remaining precursor ion peaks. The addition of step does not affect the results from our test data, but was added to ensure deletion of the precursor ions. The third step is to eliminate peaks that are obviously not mono-isotopic. This is accomplished by examining peaks in order of intensity and deleting peaks that are within 0 to 2 Da of the $m/z$ value of the peak being examined. In our test data, this step increases the number of true positive hits in the 100 fmol dataset by 69%.

The last step in the noise filter is to filter out peaks that are too close together. In this step, precursor charge 1+ and 2+ spectra are treated differently from charge 3+ spectra. It is assumed that 1+ and 2+ spectra contain mainly 1+ product ions. For these spectra, the filter examines each peak in order of intensity. In a region ±27 Da of the peak being examined, all except the most intense peak is deleted under the assumption that in a region of this size, no other peak from the same ion series will exist and only one ion from the complementary ion series can exist. 27 Da is used as it is less than the residue mass of the smallest amino acid. This value is user adjustable.

For precursor charge 3+, the spectra are assumed to contain charge 1+ product ions in the range above $m/2$ and 1+ and 2+ product ions in the region below $m/2$, where $m$ is the neutral mass of the precursor. The 1+ product region is treated using the same filter as the 1+ and 2+ precursor ions. In the 2+ product region, a slightly different filter is applied. Peaks are examined in order of intensity and only the two most intense peaks within ±14 Da of the peak being examined are kept. 14 Da is used as it is less than the residue mass of the smallest amino acid doubly charged. Two peaks are allowed to account for two types of ions: a doubly charged ion of the complementary series or a singly charge ion of either series.

Ions corresponding to water and ammonia loss from b and/or y ions are excluded in the last step by ignoring peaks that are 17 or 18 Da lower in mass than the peak being examined. Additionally, to avoid deleting monoisotopic peaks, peaks 1 Da less than the peak being examined are not considered.

In our 100 fmol test data, this fourth step increased the number of true positives hits by 16%.

**Calculation and comparison of precursor mass.** The first comparison done between the experimental spectrum and the sequence search library is to compare the measured precursor mass and those calculated by digesting *in silico* the search library with the enzyme of interest. If the masses match within a tolerance given by the user, the algorithm proceeds to the next step, otherwise it selects the next peptide from the sequence search library for precursor mass comparison. The computation of the theoretical peptide mass allows for missed cleavage, fixed modifications to the mass of an amino acid, and variable modifications to the mass of an amino acid, where "variable" means that masses are calculated with and without the modification.

This step is the most computationally intensive step in OMSSA. In order to speed the algorithm, several strategies are used. First, computation is done using integers – all masses are scaled by a factor of 100 to deal with non-integer masses. This scaling value may be adjusted by the user in the source code. Secondly, the sequence library is memory mapped for quickly loading the sequences into the CPU. Memory mapping is a technique where a file is accessed as though it were main memory. Thirdly, the spectra being searched are sorted and indexed by precursor mass, avoiding unnecessary comparisons.

**Calculation of the mass ladder.** If the precursor mass matches a calculated mass, the theoretical *m/z* values of the product ions are calculated from the search library peptide for comparison to the *m/z* values derived from the experimental spectrum. $b^{1+}$ and $y^{1+}$ ion series are calculated for charge 1+ and 2+ precursor ions and, additionally, $b^{2+}$ and $y^{2+}$ product ions are calculated for charge 3+ precursor ions. Restricting 2+ precursor ions to only producing 1+ product ions increases the number of true positive hits in the 100 fmol test set by 79%.

If variable modifications are selected by the user, multiple ladders are calculated for each theoretical peptide. The ladders calculated in OMSSA are monoisotopic. In the mass range normally used in high throughput proteomics, this should not cause a significant loss in sensitivity, although in future versions of the algorithm we will address the issue.

**Comparison of Mass Ladders.** To find hits, OMSSA compares the calculated mass ladders with the mass ladders derived from the experimental spectra using a mass tolerance given by the user. Both mass ladders are sorted by *m/z* value for fast matching.

If an experimental *m/z* value is used for a match between a given experimental mass ladder and a calculated mass ladder, it is no longer considered for matching ions in other ion series, e.g. if an experimental *m/z* value matches a particular b ion, it is not allowed to match a y ion. The reason for this restriction is that if it is not possible to tell within an instrument's resolution if particular *m/z* value matches more than one ion species, then a conservative approach is to assume that it matches only one. If one assumes otherwise, the false positive rate increases due to peptides in the sequence library that have a large number of matching ion species, for example, palindromic peptides.

The number of matched ions is inserted into a list ranked by the number of matches. There is a separate list for each charge state the precursor may be in. To conserve memory, the length of this ranked list is limited to a user adjustable value, 100 by default.

**Scoring.** To determine a scoring function, it is useful to understand the characteristics of random matches to *m/z* values derived from spectra. Calculating a distribution of random matches allows the significance of a hit to be expressed as the probability of the hit being random, where a low probability implies a significant hit. To study the distribution, a randomly selected set of ion trap spectra from an LC-MS/MS study of tryptically digested yeast whole cell lysate was searched against the NCBI non-redundant protein sequence library (nr). Figure 2 shows a histogram for one of these spectra that counts the number of product *m/z* values in each theoretical peptide that match the product *m/z* values of the experimental spectrum when the precursor mass of the experimental spectrum matches the calculated precursor mass. Fitted to the histogram is a Poisson distribution, which is a distribution found in random processes where the average number of successes is much less than the possible number of successes. This relationship to the Poisson distribution has been described in earlier studies [23, 27]. The good fit to the Poisson distribution motivated us to devise a model of the random hit distribution that does not rely on peak fitting or iterating through a sequence library for its calculation.

To create the model, we first consider spectra that contain only charge 1+ product ions. Let *o* be the lower bound of measured product ions *m/z* values and *r* be the upper bound. If the product ion measurement tolerance is *t*, then a measure of the number of possible matches is (*r-o*)/2*t*. If *m* is the neutral mass of the precursor, we are trying to match *h·*(*r-o*)/*m* calculated *m/z* values to *v* experimental product ions, where *h* is the total number of calculated *m/z* values for product ions. Assuming a Poisson process, this would give us a mean

$$\mu_1 = \left(\frac{2t}{(r-o)}\right) \cdot \left(\frac{h \cdot (r-o)}{m}\right) \cdot v = \frac{2thv}{m}$$

for the Poisson distribution

$$P(x, \mu) = \frac{\mu^x}{x!} e^{-\mu}$$

where *x* is the number of matches measured.

Next, we consider spectra that contain 1+ and 2+ product ions. The probability distribution is also a Poisson distribution whose mean is given by

$$\mu_2 = \mu_1 \cdot \frac{r + m - 3o}{r - o}$$

The details of this derivation are given the Appendix. Note that in the model, the effect of random noise is taken into account by including the number of noise peaks in $v$, allowing OMSSA to use a noise filter that does not filter all noise peaks.

To help validate the model, nine spectra from the yeast lysate set were selected with charge states of 1+, 2+ and 3+ and neutral masses ranging from 1090 Da to 3570 Da. Each spectrum was compared to all theoretical peptides calculated from the nr search library with matching precursor ion masses and charge, and the number of product ion matches recorded. The mean value of the number of matches was measured from this set. The model mean was calculated using the average value of the model for the top 100 hits and plotted against the measured mean in Figure 3. Note that the correspondence is approximately one-to-one, indicating that the calculated Poisson distribution outlined above may be a sufficient measure of the random distribution of hits to the *m/z* values of a spectra. To further explore the relationship of the model and measured mean with respect to the product mass tolerance $t$, the tolerance was doubled from 0.8 Da to 1.6 Da. As shown in Figure 3, the linear relationship still holds.

**Selection of *m/z* values matching the most intense peaks.** To make the algorithm more efficient and increase sensitivity, the following selection on the theoretical spectra is made: At least one of the *m/z* values of the theoretical spectra must match the *m/z* values of the top *n* peaks in the spectrum ($n=3$ by default). This selection changes the probability distribution. If the probability $q$ that a matched *m/z* value matches a calculated *m/z* value is $n/v$, then the probability distribution is

$$P'(x,\mu) = \frac{1}{Q}\left(1-(1-q)^x\right)P(x,\mu)$$

where the normalization factor $Q$ is

$$Q = \sum_x \left(1-(1-q)^x\right)P(x,\mu)$$

over all $x$, and can be computed numerically.

**E-value calculation.** OMSSA reports hits ranked by E-value. An E-value for a hit is a score that is the expected number of random hits from a search library to a given spectrum such that the random hits have an equal or better score than the hit. For example, a hit with an E-value of 1.0 implies that one hit with a score equal to or better than the hit being scored would be expected *at random* from a sequence library search. If the probability that a single comparison of one spectrum to one calculated ms/ms spectrum is not random is

$$\sum_{x=0}^{y-1} P'(x,\mu_z)$$

where $y$ is the number of successful product ion matches and $z$ is 1 or 2 depending on the ion series searched, then the probability that a search of one spectrum against $N$ theoretical spectra is random is

$$1 - \left(\sum_{x=0}^{y-1} P'(x,\mu_z)\right)^N$$

The E-value is then

$$E(y,\mu) = N\left(1 - \left(\sum_{x=0}^{y-1} P'(x,\mu_z)\right)^N\right)$$

This E-value remains valid for searches that include variable post-translational modifications. In general, variable post-translational modifications increase the value of $N$ since modified peptides can generate two or more theoretical spectra, depending on the number of modified sites.  The multiple theoretical spectra generated from a single peptide sequence can be considered non-redundant, as they do not have the same precursor *m/z* and only share a subset of the product ions, making it unnecessary to explore the effect of redundancy on the E-value.

**Rescoring to improve sensitivity.** The sensitivity of OMSSA can by improved by varying the threshold used in the initial step of the noise filter to cut off background noise.  This is accomplished by varying the background noise threshold from 0-20% of the maximum intensity peak and examining the E-value of the best hit.  The final threshold chosen is the one the results in the lowest E-value for the best hit.   It is possible that this rescoring could affect the statistics adversely by resampling the sequence library, however, in practice this does not significantly change the list of best hits.

**OMSSA validation and comparison to Mascot.** Matrix Science's Mascot includes a MS/MS search algorithm with probability-based scoring, so it is useful to validate OMSSA by comparison to Mascot.  Both OMSSA and Mascot were used to analyze 486 spectra from a 100 fmol protein standard and 134 spectra from a 10 fmol protein standard.  The two different sets of spectra were used to ensure a range of spectrum quality.

For every hit to a spectrum, Mascot reports an ions score and two score thresholds – the identity threshold and the homology threshold.  Although the exact details of the Mascot scoring statistics have not been published, an ions score higher than the identity threshold is defined to mean that the hit has a less than 5% probability of being a random event.   In this paper we report Mascot results ranked by the ions score minus the identity threshold.  We do not use the homology threshold as this did not appear to improve the overall Mascot results.

Since the nr sequence library included multiple homologous forms of the proteins found in the standard cocktail, we performed a two step sequence comparison between the hit peptides and the true peptides to determine if the hits were true positives. The first comparison was a residue by residue comparison where the residue pairs isoleucine/leucine and lysine/glutamine were considered the same within the resolution of the ion-trap mass spectrometer. If this comparison succeeded, the hit was labeled a true positive. If the first comparison did not succeed and the peptide was longer than 5 residues, pair wise BLAST was performed between the hit peptide and the true peptides. If a BLAST hit with a BLAST e-value below 0.02 was found, the hit was labeled homologous but not identical and eliminated from the analysis, i.e. it was considered neither a true nor a false positive. The reason for this rejection is that OMSSA was designed to be an identification algorithm and does not explicitly take into account homology. All remaining hits with e-values above 0.02 were counted as false positives.

Table 1 lists significant top hits to spectra as identified by OMSSA and Mascot. The hits are categorized into true and false positives. A significant hit in OMSSA has an E-value < 0.1 and a significant hit in Mascot has an ions score that exceeds the identity threshold. For both 10 fmol and 100 fmol protein standards, OMSSA identifies a larger number of spectra than Mascot. Neither algorithm has a top hit above significance threshold that is a false positive.

The search result scores for all hits found by searching the 100 fmol and 10 fmol protein standards are shown as categorized histograms in Figure 4. The OMSSA significance threshold correctly categorizes true and false positive hits. The two significant Mascot false positive hits correspond to the peptide SSLRQTVVR from the human protein GREB1, isoform a (accession NP_055483). Neither of these significant false positives is the top scoring hit -- Mascot finds a better scoring hit to chicken lysozyme in both cases. A BLAST search of this peptide to all mammalian and chicken sequences in nr show no significant homologs to the cocktail proteins or obvious contaminants, making this a likely true false positive. This peptide has several closely matching b and y ions of charge 1+ and 2+, which can lead to over counting of ion matches as discussed earlier in this paper.

Commonly, Receiver Operating Characteristic (ROC) analysis is applied to characterize the sensitivity and specificity search algorithms [28]. ROC curves for OMSSA and Mascot are show in Figure 5. For the 10 fmol data set, OMSSA displays slightly better specificity than Mascot and slightly worse sensitivity. For the 100 fmol data set, the sensitivity and specificity of OMSSA is better than Mascot. The differences between these two ROC analyses are likely due to a variety of factors, including variation in signal to noise between the two datasets.

Although we did not perform extensive performance testing, OMSSA took 901 seconds on one CPU to perform the 100 fmol search. Mascot required 716 seconds running on a cluster of 6 roughly equivalent computational CPUs.

**Conclusion**

The ROC analysis shows that, within the limitations of the spectral dataset, OMSSA is an efficient, sensitive, and specific algorithm for matching MS/MS peptide spectra. At default thresholds, OMSSA matches more spectra from a standard protein cocktail than Mascot and is faster than Mascot in searching large datasets.

**Appendix**

Derivation of the Poisson distribution for spectra containing doubly and singly charged product ions. If 1+ and 2+ product ions are present in an MS/MS spectrum, the spectrum can be considered as two separate ranges: A, the *m/z* range above *m*/2, which contains only charge 1+

product ions and B, the *m/z* range below *m/2*, which contains charge 1+ and 2+ product ions. Each of these regions can be modeled with a separate Poisson distribution. To begin, we will consider range A. For this range, the number of possible matches is (*r-m/2*)/2*t* and we are trying to match *h*·(*r-m/2*)/*m* calculated *m/z* values to *v*·(*r-m/2*)/(*r-o*) experimental product ions. This would give us a mean of

$$\mu_A = \left(\frac{2t}{r-m/2}\right)\left(\frac{h\cdot(r-m/2)}{m}\right)\left(\frac{v\cdot(r-m/2)}{r-o}\right) = \frac{2thv}{m}\cdot\frac{r-m/2}{r-o}$$

Range B has (*m/2-o*)/2*t* possible matches. In this range we are trying to match *h*·(*m/2-o*)/*m* singly charged ions and *h*·(*m/2-o*)/(*m/2*) doubly charged ions to *v*·(*m/2-o*)/(*r-o*) experimental product ions. The resulting mean is

$$\mu_B = \left(\frac{2t}{m/2-o}\right)\left(\frac{h\cdot(m/2-o)}{m} + \frac{h\cdot(m/2-o)}{m/2}\right)\left(\frac{v\cdot(m/2-o)}{r-o}\right) = \frac{6thv}{m}\cdot\frac{m/2-o}{r-o}$$

From elementary probability theory, the mean of the combined distribution is

$$\mu_2 = \mu_A + \mu_B = \frac{2thv}{m}\cdot\frac{r+m-3o}{r-o}$$

## Acknowledgment

We thank David Lipman, John Spouge, and Renata Geer for helpful discussions, and Jeri Roth for assistance in obtaining the test spectra. We greatly appreciate the programming assistance of the NCBI Information Engineering Branch headed by Jim Ostell. We are also grateful to the NIH intramural research program for support.

Table 1. Numbers of spectra identified in the 10 fmol and 100 fmol data sets, categorized by the highest scoring hit above significance threshold. true pos: true positive, false pos: false positive.

|  | 10 fmol | | 100 fmol | |
| --- | --- | --- | --- | --- |
| Algorithm | true pos | false pos | true pos | false pos |
| OMSSA | 27 | 0 | 100 | 0 |
| Mascot | 24 | 0 | 73 | 0 |

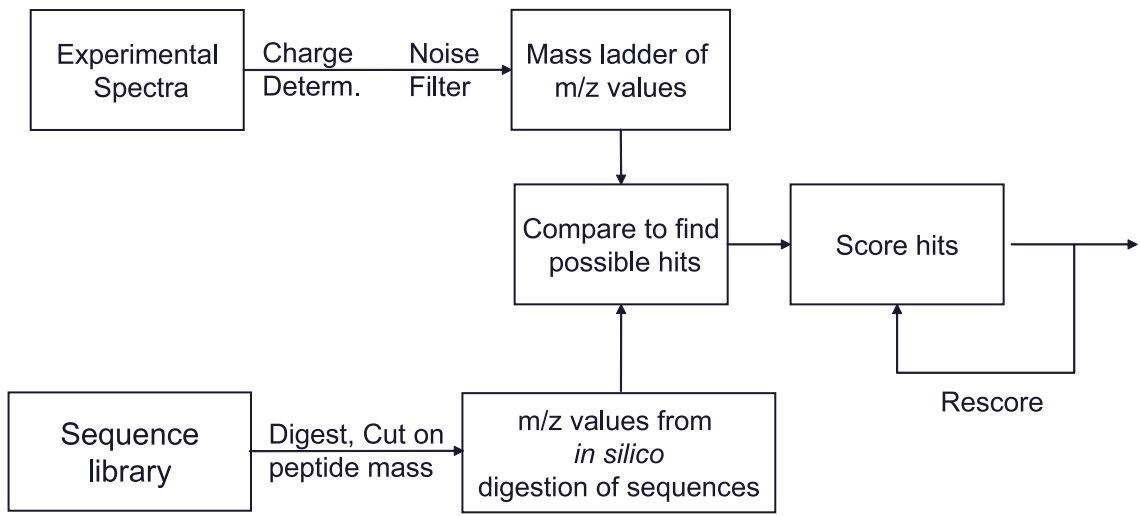

**Figure 1.** Algorithm flow diagram.

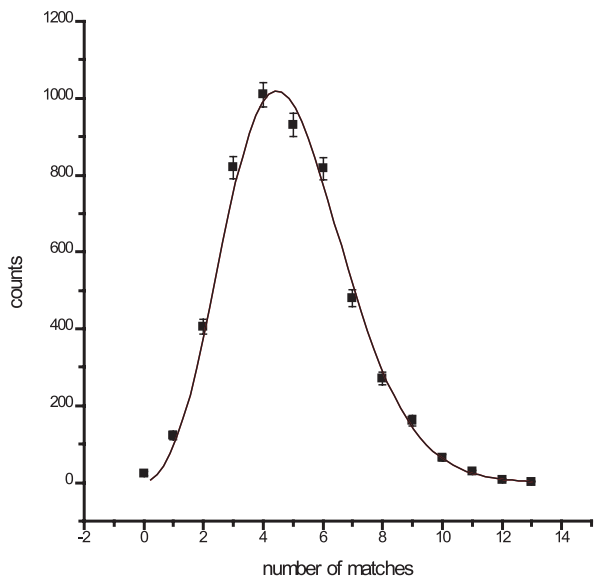

**Figure 2.** Histogram of the number of product ion matches for a yeast lysate spectrum against all peptides in the NCBI nr sequence library whose precursor mass matches the lysate spectrum precursor mass. Error bars are the square root of the counts. The fit curve is a Poisson distribution.

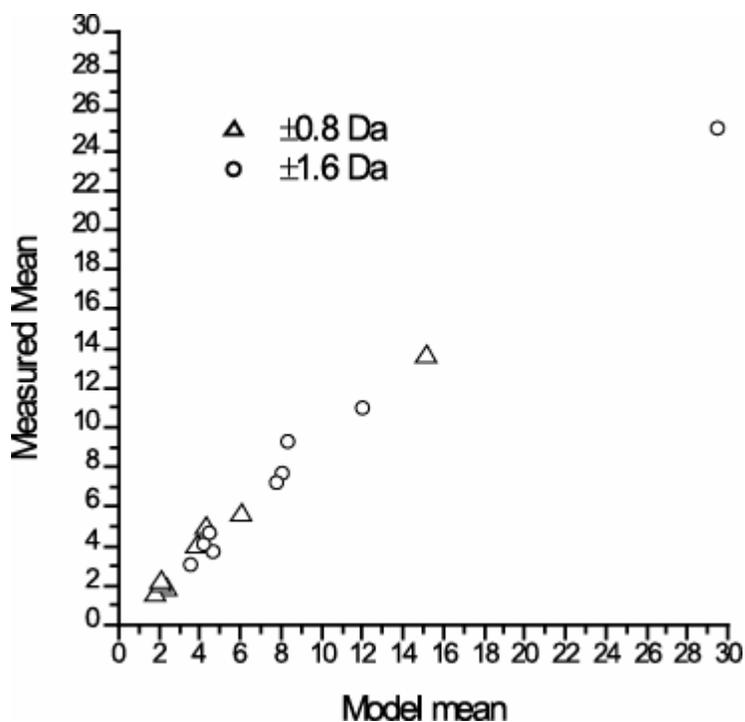

**Figure 3.** Comparison of the mean values of the number of matches of theoretical *m/z* values to experimental data and the mean values calculated using the model. Each point corresponds to a different MS/MS spectrum searched against the nr sequence library. The triangles correspond to searches performed with a product ion tolerance of ±0.8 Da and the circles correspond to a product ion tolerance of ±1.6 Da.

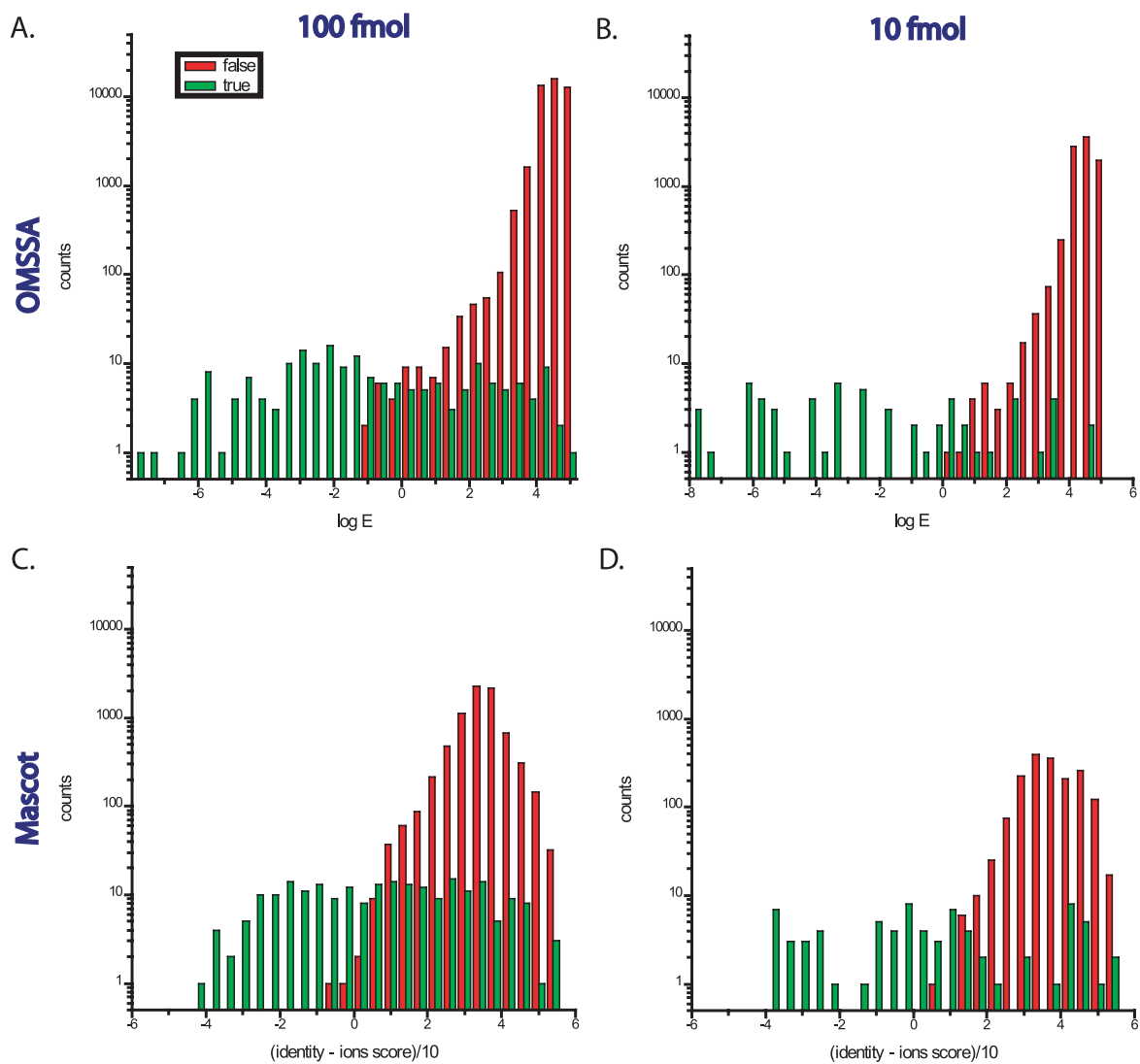

**Figure 4.** Score histograms categorized into true and false positives. A and B are for the 100 fmol and 10 fmol protein standards, respectively, as analyzed by OMSSA. Red indicates false positives and green true positives. The horizontal axis is the logarithm of the E-value and the vertical axis the number of hits with the given score. Scores below an E-value of 0.1 are considered significant. C and D are for the 100 fmol and 10 fmol protein standards as analyzed by Mascot. The horizontal axis is the identity threshold minus the ions score divided by 10, and values below 0 are considered significant. This scale was chosen to allow comparison to the OMSSA score.

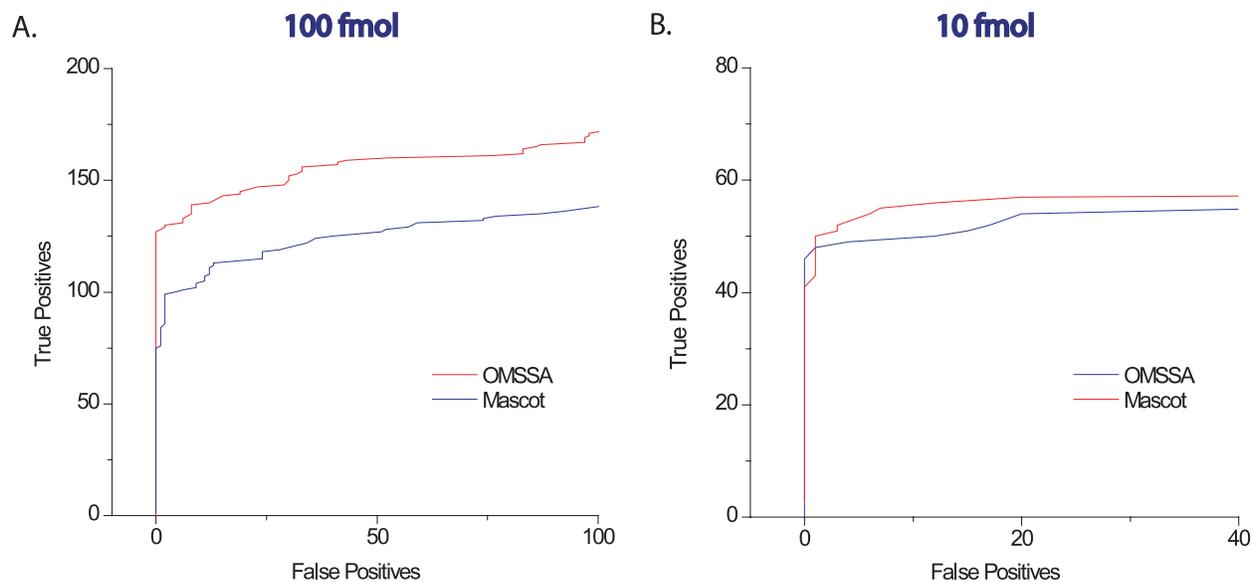

**Figure 5.** ROC analysis plots for the 100 fmol and 10 fmol protein standards as searched by OMSSA and Mascot. The values plotted are the total number of true and false positive identifications as the significance threshold for each algorithm is varied from high significance to low significance. A value towards the top indicates higher sensitivity and a value to the left indicates higher specificity.

TOC synopsis: An efficient MS/MS search algorithm with an explicit probability score developed with classical hypothesis testing

TOC graphic:

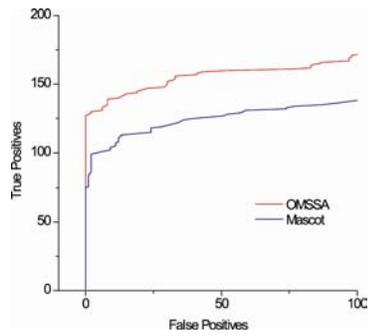